\documentclass[prl,twocolumn]{revtex4}
\usepackage{dcolumn}% Align table columns on decimal point
\usepackage{bm}% bold math

\usepackage{graphics}
\usepackage{graphicx}
\usepackage[dvips,usenames]{color}

\newcommand{\vs}{\quad}

\newcommand{\be}{\begin{equation}}
\newcommand{\ee}{\end{equation}}
\newcommand{\ba}{\begin{eqnarray}}
\newcommand{\ea}{\end{eqnarray}}
\newcommand{\NL}{\nonumber \\}

\newcommand{\eqref}[1]{(\ref{#1})}
\renewcommand{\d}{\partial}

\newcommand{\exv}[1]{\left\langle{#1}\right\rangle}

\renewcommand{\c}[1]{{\cal{#1}}}

\begin{document}

%%%%%%%%%%%%%%%%%%%%%%%%% TITLE PAGE %%%%%%%%%%%%%%%%%%

\title{Power-law tails from multiplicative noise}
%\title{Power-law tails from a modified Brownian motion}
%\title{Power-law tails and noisy equations}
%\title{Power-law tails are omnipresent}

\author{Tam\'as S. Bir\'o}
\affiliation{MTA KFKI Research Institute for
Particle and Nuclear Physics, H-1525 Budapest Pf. 49, Hungary}

\author{Antal Jakov\'ac}
\affiliation{Research Group for Theoretical Condensed Matter of HAS and
TU Budapest, H-1521 Budapest, Hungary}

%\date{ \today}

\begin{abstract}
\vs
{\bf Abstract}
We show that the well-known linear Langevin equation, modeling the Brownian
motion and leading to a Gaussian stationary distribution of the corresponding
Fokker-Planck equation, is changed by the smallest multiplicative noise.
This leads to a power-law tail of the distribution for large enough
momenta. At finite ratio of the correlation strength for the
multiplicative and additive noise the stationary energy distribution
becomes exactly the Tsallis distribution.
\end{abstract}


\maketitle

%%%%%%%%%%%%%%%%%%%%%%%%%  MAIN  TEXT BODY %%%%%%%%%%%%%%%%%%%%%%%

%\section{Introduction}

%\vs
Power-law tails are present in numerous distributions studied in physics
or elsewhere when dealing with complex systems\cite{POWER-TAILS}. 
They are of generic
interest, regarded as to signal long range order, non-vanishing
correlations or scale invariance in complex systems with strong
dynamics, which's details are mostly unknown.%\cite{LONG-RANGE}. 
These are often contrasted
to the traditional statistical system, showing the Gibbs distribution
($\exp(-E/T)$) in energy, which is Gaussian in the momenta of
free, massive particles ($\exp(-p^2/2mT)$). The latter is considered
as the generic case for thermal equilibrium of non-correlated or short-range
correlated systems. This concept has been carried far beyond of its
original field describing monoatomic ideal gas (Maxwell-Boltzmann statistics),
by applying the Gibbs distribution in thermal equilibrium
to areas such as particle physics and field theory. It serves as starting
point of high-temperature field theory calculations both with analytical
and numerical methods. Lattice gauge theory is based on the formal
similarity between Euclidian path integrals and the canonical partition
sum.

%\vs
A very simple and elegant, microdynamical explanation for the Maxwell-Boltzmann
statistics is offered by the Langevin equation, describing a free
particle moving under the influence of a deterministic damping force
and a stochastic drive\cite{LANGEVIN}. 
The latter accelerates the particle in a short
time, changing its momenta randomly and uncorrelated. The stationary solution of
this stochastic equation follows the Gaussian statistics, compatible
to Gibbs' principle. It seems that in many statistical considerations of 
complex physical models from that on it is tacitly assumed that the
presence of this additive noise is a dominant effect: the equilibrium
distribution follows Gibbs' formula (the Gaussian distribution in momentum
for a free, massive particle). Since the harmonic oscillator is just the
extension of this free motion Langevin equation into the phase space,
also for free quantum systems the above picture is generally accepted.

%\vs
We shared this expectations, thinking that any non-Gaussian (or in the
energy non-exponential) distribution, especially
the power-law tail observed in many phenomena including particle
spectra in high energy physics, may only come from non-thermalized or
in an other way non-equilibrium situation. In this note we would like
to share our deep astonishment about that this is not so: we found out
that treating the damping constant in the Langevin equation also
stochastically (considering this way both multiplicative and additive noise)
the stationary distribution is in general non-Gaussian.
Moreover, above a certain momentum, depending on the strength
(i.e. self-correlation width) of the multiplicative noise, the
stationary solution goes over into a power-law.

%\vs
This finding seems to have manifold consequences. Even in equilibrium,
even averaging over a huge number of elementary events, measurements
on quantum systems, such as particles or fields, are bound to find
a power-law tail irrespective if the underlying dynamical system
had equilibrated itself long enough in units of a characteristic time.
Power-law tails in pion spectra, found experimentally in  high energy $e^+e^-$,
$p\overline{p}$ or heavy-ion collisions\cite{POW-TAIL-SPECTRA}, 
in the view of this simple
mathematical result may reflect an already stationary distribution.
Power-law tails found in other areas, in particular as properties of
average correlations (cf. time series in stock market), may also be
a non-temporary, long term effect.

%\vs
In this paper we derive a generic stationary distribution for the
Langevin-type equation with both additive and multiplicative noise.
Conform to the original assumptions both noise terms are white
(Dirac delta correlated in time), but they may show cross-correlations.
This situation may stem naturally from the widespread treatment
of field theoretical operator equations, where the fields are split to 
a large, ''classical'' expectation value and to a noisy quantum or
thermal fluctuation part: the different noise terms do have a common
origin, so it is natural to consider cross-correlations among them
\cite{NOISE-FIELD}.
Often the overdamped approximation is applied for studies of plasmas,
solving then effectively a differential equation first order in time,
instead of second order\cite{PLASMA}. We think therefore that the analytically
solvable case, we present in this paper, may carry a quite general lesson.

%\begin{document}
%\newpage
%\section{Langevin equation with two noise terms}
%\label{sec:LangFock}

We present the solution of the Langevin equation
with both additive and multiplicative noise terms applying the
classical method of Wang and Uhlenbeck\cite{UHLENBECK}. Our starting point is
the linear equation:
\begin{equation}
  \label{eq:lan}
  \dot p + \gamma p = \xi,
\end{equation}
where now both $\xi$ and $\gamma$ are  stochastic variables.
They both may have a non-zero mean value (motivated by possible field theory
applications),
\be
\exv{\xi(t)}=F, \qquad \exv{\gamma(t)}=G,
\label{MEAN}
\ee
and show white-noise (i.e. extremely short term) correlations:
\ba
\exv{\xi(t)\xi(t')} -\exv{\xi(t)}\exv{\xi(t')}\: &=& \: 2D \: \delta(t-t'),   \NL
\exv{\gamma(t)\gamma(t')} -\exv{\gamma(t)}\exv{\gamma(t')}\: &=& 
	\: 2C \: \delta(t-t'), \NL 
\exv{\gamma(t)\xi(t')} -\exv{\gamma(t)}\exv{\xi(t')} \: &=& \: 2B \: \delta(t-t'), \NL
\exv{\xi(t)\gamma(t')} -\exv{\xi(t)}\exv{\gamma(t')} \: &=& \: 2B \: \delta(t-t'),
\label{CORR}
\ea
This problem can be solved analytically.
We will determine the time dependence of the distribution of $p$
values, denoted by $f(p,t)$. In this notation $f(p_0,t)dp$ is the
probability that after time $t$ the variable $p$ has the value in the
range $[p_0,p_0+dp]$. We rewrite \eqref{eq:lan} as a difference
equation
\begin{equation}
  \label{eq:lan1_diff}
  p(t+dt) = p(t)  \: + \:
  \int\limits_t^{t+dt}\!dt'\,\left(\xi(t') - \gamma(t')\: p(t')\right), 
\end{equation}
If $p(t)$ is a smooth function of time we can
replace $p(t')$ either by $p(t)$ or $p(t+dt)$, or any value in between.
We choose here the Ito prescription\cite{ITO}, 
which uses $p(t)$ under the integral in the $dt \rightarrow 0$ limit.
In order to simplify notation we
write the integral term as $dt\exv{x}$, with $x$ denoting the general
integrand.
Now we can write down a Fokker-Planck equation for the distribution:
the probability to have the value \hbox{$p(t+dt)$} at $t+dt$ is the probability
that we have the value $p(t)$ at time $t$ and noise values $\xi$ 
and $\gamma$ that just satisfy \eqref{eq:lan1_diff}:
\begin{equation}
  \label{eq:FP_pro}
  f(p,t+dt) = \int\!d\xi d\gamma\, P(\xi,\gamma)\, 
  		f(p - dt \exv{\xi}+p dt \exv{\gamma}).
\end{equation}
Unfortunately this form is not appropriate to directly create
differential equation as $dt\to0$. Instead we follow the method of
Wang and Uhlenbeck: we have a trial function $R(p)$ that is smooth
enough and we compute the expectation value of $R$ (averaged over the
noise) as a function of time
\begin{equation}
  \exv{R(t)} = \int\!dp\, R(p)\, f(p,t).
\end{equation}
Applying this form to \eqref{eq:FP_pro} we have 
\ba
  \int\!dp\, R(p)\, f(p,t+dt) = \qquad \qquad \qquad  & &\NL 
  \int\!d\xi d\gamma\,P(\xi,\gamma)
  \int\!dp\, R(p + dt\: \exv{\xi} - p \:dt\: \exv{\gamma} )\, f(p,t). & &
\ea
%We now assume that the two variables are independent, so
%$P(\xi,\gamma)$ factorizes. 
By Taylor expanding $R(p)$  and integrating over the noise distribution we get
\ba
  \exv{ R(p + dt\: \exv{\xi} - p \:dt\: \exv{\gamma} ) } =  \qquad \qquad \qquad & & \NL  \NL
  \exv{R(p) + dt \,
  R'(p) \: K_1(p) + dt \, R''(p) \: K_2(p) + \c O(dt^2)} & &
\ea
with
\be
K_1 = F - Gp, \qquad  K_2 = D - 2B p + Cp^2
\label{K1K2DEF}
\ee
in the present case.
This leads to the following general Fokker-Planck equation
\begin{equation}
  \label{eq:FP_gen}
  \frac{\d f}{\d t} = -\frac{\d (K_1 f)}{\d p} +
  \frac{\d^2\!(K_2 f)}{\d p^2}.   
\end{equation}
The stationary solution satisfies 
\begin{equation}
  \frac{d}{dp} \: (K_2f) \, = \, K_1 \: f,
\end{equation}
which is analytically solvable. It leads to
\be
f(p) \, = \, f(0)\: \frac{K_2(0)}{K_2(p)}\: \exp(\:L(p)\:)
\label{FP-stac}
\ee
with
\be
L(p) \: = \: \int_0^p\!\!dq \, \frac{K_1(q)}{K_2(q)}.
\ee
In the case of two noises correlated the way given in eq.\eqref{CORR}
we arrive at the following logarithm of the stationary
distribution:
\be
\ln \frac{f(p)}{f(0)} = -\left(1+\frac{G}{2C}\right)\ln \frac{K_2(p)}{D}
\: - \: \frac{\alpha}{\vartheta} \: {\rm atn} \left( \frac{\vartheta \: p}{D-Bp} \right),
\label{SOLUTION}
\ee
with
\be
\vartheta = \sqrt{CD-B^2} \qquad {\rm and} \qquad \alpha = G \: \frac{B}{C}-F.
\ee
Here 'atn' denotes the inverse tangent function, not always taking the 
first principle value, but rather continuing at $p>D/B$ smoothly.
%(This can be achieved in programming by using the $atan2(x,y)$ function.)
Considering physical applications the noise correlation values,
$D$, $C$ and $B$ build a positive semidefinite matrix. This ensures
that $C$ and $D$ are non-negative values, and the determinant
$\vartheta$ is real and also non-negative. The same applies for
the function $K_2(p)$ occurring under the logarithm.
In this context zero
values are limiting cases and the stability of the stationary
solution \eqref{SOLUTION} against choosing a small
positive value has to be investigated.

%\vs
The lesson for physical applications lies in the analysis of different
limiting cases. First we consider the traditional case: $C=B=0$,
not allowing for any noise (fluctuation) in the multiplicative factor
$\gamma$ (damping constant). This leads back to the familiar
Gauss distribution:
\be
f(p) = f(0) e^{-\frac{G}{2D} p^2} e^{\frac{F}{D}p}
\label{GAUSS}
\ee
with an eventual shift in the mean momentum for a non-zero mean
driving force $\exv{\xi}=F$.
Another limiting case is that of the purely multiplicative noise
with $D=0, B=0$
(G.Wilk\cite{WILK} has applied it for the heat conduction equation and obtained
a Gamma distribution for the inverse temperature $1/T$). Now the
stationary solution becomes
\be
 f(p) = f(0) \: p^{-\: 2\: -\: G/C} \: e^{-\frac{F}{C \: |p|}}
\label{GAMMA}
\ee
a Gamma distribution in $1/p$. For large $p$ this approaches a
pure power-law.
It is particularly interesting to investigate the mathematically degenerate
case of $\vartheta=0$. Now $K_2(p)$ reaches zero at the critical
momentum, $p_c=\sqrt{D/C}$, leading to zero probability in the
stationary distribution $f(\sqrt{D/C})=0$. This occurs as a
''limiting momentum'' in the physical distribution.

\begin{figure}
\begin{center}
\includegraphics[width=0.25\textwidth,angle=-90]{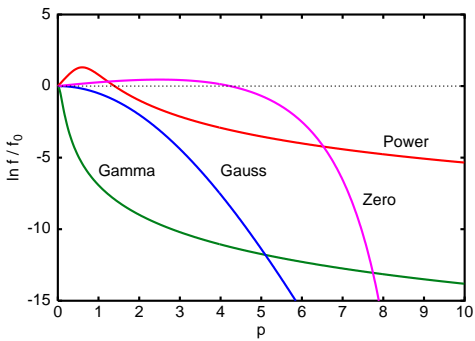}
\end{center}
\caption{\label{LOG-f-CASES}
 Comparison of the generic stationary distributions of the Langevin-type
 equation with both additive and multiplicative noise. Limiting cases,
 as the Gauss-, the Gamma- and the Power-Law distribution are labeled
 correspondingly. The label 'Zero' refers to the case with $\vartheta=0$.
% The numbers in the legend code the parameters applied
% in the order $G, F, D, C,$ and $B$. $s$ stands for a small number,
% $L$ for a large one. The critical momenta in these cases are
% $1$ GeV for the Power-Law, $0.1$ GeV for the Gamma-, $10$ GeV
% for the Gauss- and for the degenerate 'Zero' distribution.
   }
\end{figure}

%\vs
The above limiting cases rely on an expansion of the
generic solution. They are valid only in a limited range
of momenta: the Gaussian solution for  small, 
the Gamma distribution for large argument of the inverse tangent
function. Correspondingly the widely beloved Gaussian
distribution can be a good approximation
to the stationary solution only for momenta $p \ll \sqrt{D/C}$.
(For $C=0$ strictly, of course this is 'the' solution for any finite momentum.)
Generally the small argument of the inverse tangent is fulfilled
for
%\be
$p \ll \frac{D}{\vartheta + B} $.
%\label{KNACK}
%\ee
This result also means that {\em for the smallest fluctuation
in the multiplicative factor the stationary Gauss distribution
develops a power-law tail}. The power in this tail, $p^{-2v}$, is given by 
%\be
$v = 1+\frac{G}{2C}$.
%\label{POWER}
%\ee
In order to offer a visual insight into the nature of
the generic stationary solution we show stationary spectra
for different parameters (cf.Fig(\ref{LOG-f-CASES})).

%%%%%%%%%%%%%%%%%%% FIG:  log distributions, case analysis
%\newpage
%\vspace*{-20mm}

%%%%%%%%%%%%%%%%%%%%%%%%%%%%%%%%%%%%%%%%%%%%%%%%%%%%%%%%%%%%%%%%

%\vs
In the case $B=0$ (no cross correlation between the noises), and
$F=0$ (no drift term due to the additive noise), the {\em exact} stationary
distribution is the Tsallis distribution\cite{TSALLIS-DIST}. 
In order to achieve this
result one uses the energy of the free particle, $E=p^2/2m$ as the
distribution variable and eq.\eqref{SOLUTION}. We get
\be
 f(E) = f_0 \, \left(1 + (q-1) \frac{E}{T} \right)^{\frac{q}{1-q}}
 \label{TSALLIS}
\ee
where the parameters of the Tsallis distribution are given by
\be
  T = \frac{D}{mG}, \qquad q = 1 + \frac{2C}{G}.
\label{TS-PARAMS}
\ee
Again for $C=0$ (only additive noise) $q=1$ and the Tsallis distribution
goes over into the Gibbs distribution,
\be
 f(E) = f_0 \exp(-E/T). 
\label{GIBBS}
\ee
Tsallis and others have worked out a thermodynamical framework
offering the distribution \eqref{TSALLIS} as the canonical distribution.
This approach, the non-extensive thermodynamics, however, is based on
a non-extensive entropy measure. This eventually unwanted property is
not fatal: the distribution \eqref{TSALLIS} can also be obtained
based on the extensive R\'enyi entropy\cite{RENYI}. 
We note by passing this point that the presence of the two uncorrelated
noise and the corresponding Fokker-Planck equation also can be obtained
from an inhomogeneous diffusion coefficient. Instead of \eqref{eq:lan} one
may consider
\be
 \dot{p} + Gp = (D+Cp^2)^{1/2} \eta
\ee
With a single noise $\eta$, normalized to unity:
\be
 \exv{\eta(t)} = 0, \qquad \exv{\eta(t)\eta(t')}= 2 \delta(t-t').
\ee
This is a particular case of a more general field-dependent noise
considered by Arnold, Son and Yaffe in the context of non-abelian
plasmas\cite{ARNOLD}.

%\vs
The mechanism outlined in the present article may work for
non-abelian gauge theories, as well. 
%Here we just would like to sketch
%the principal mechanism and check order of magnitude estimates,
%detailed calculations are postponed for future work.
The basic gluonic field, described by a vector potential $A$, satisfies
the equation
\be
 \ddot{A} + \sigma \dot{A} = D \times (D \times A),
\label{BODEK}
\ee
with $D$ being the gauge-covariant derivative and $\sigma$ the
%inverse ohmic resistance or 
color conductivity factor. % Following Bodeker 
We consider an overdamped dynamics, when the second
time derivative is ignored. 
%This means we deal with the low-frequency behavior, we select out the soft
%fields to observe. 
Driving this to the extreme we single out
in a Fourier expansion the zero mode, $A_0$
and consider the following effective equation
\be
 \sigma \dot{A}_0 = \sum_{k,q} (k-gA_k)(q-gA_q)A_{-k-q}.
\label{PLASMA0}
\ee
Treating the the $k=0$ and $q=0$ contributions on the right hand side
separately we arrive at
\be
 \dot{A}_0 - \frac{g^2}{\sigma} A_0^3 = - \gamma A_0  + \xi,
\label{PLASM}
\ee
with
\ba
 \gamma \quad &=& \quad - 2\frac{g}{\sigma} \sum_{k>0} (k-gA_k)A_{-k},  \NL
 \xi \quad &=& \quad \frac{1}{\sigma} \sum_{k>0,q>0} (k-gA_k)(q-gA_q)A_{-k-q}.
\label{PL-XI}
\ea
Ignoring the classical $ \sim A_0^3$ contribution for the following
discussion, one realizes that $\gamma$ and $\xi$ do contain hard
Fourier component contributions. In the Langevin equation with
both additive and multiplicative noise these are regarded as noisy,
fast-fluctuating quantities.

%%%%%%%%%%%%%%%%%%% FIG:  graphical notation
%\newpage
%\vspace*{-20mm}
\begin{figure}
%\vspace*{-36mm}
\begin{center}
\includegraphics[width=0.25\textwidth]{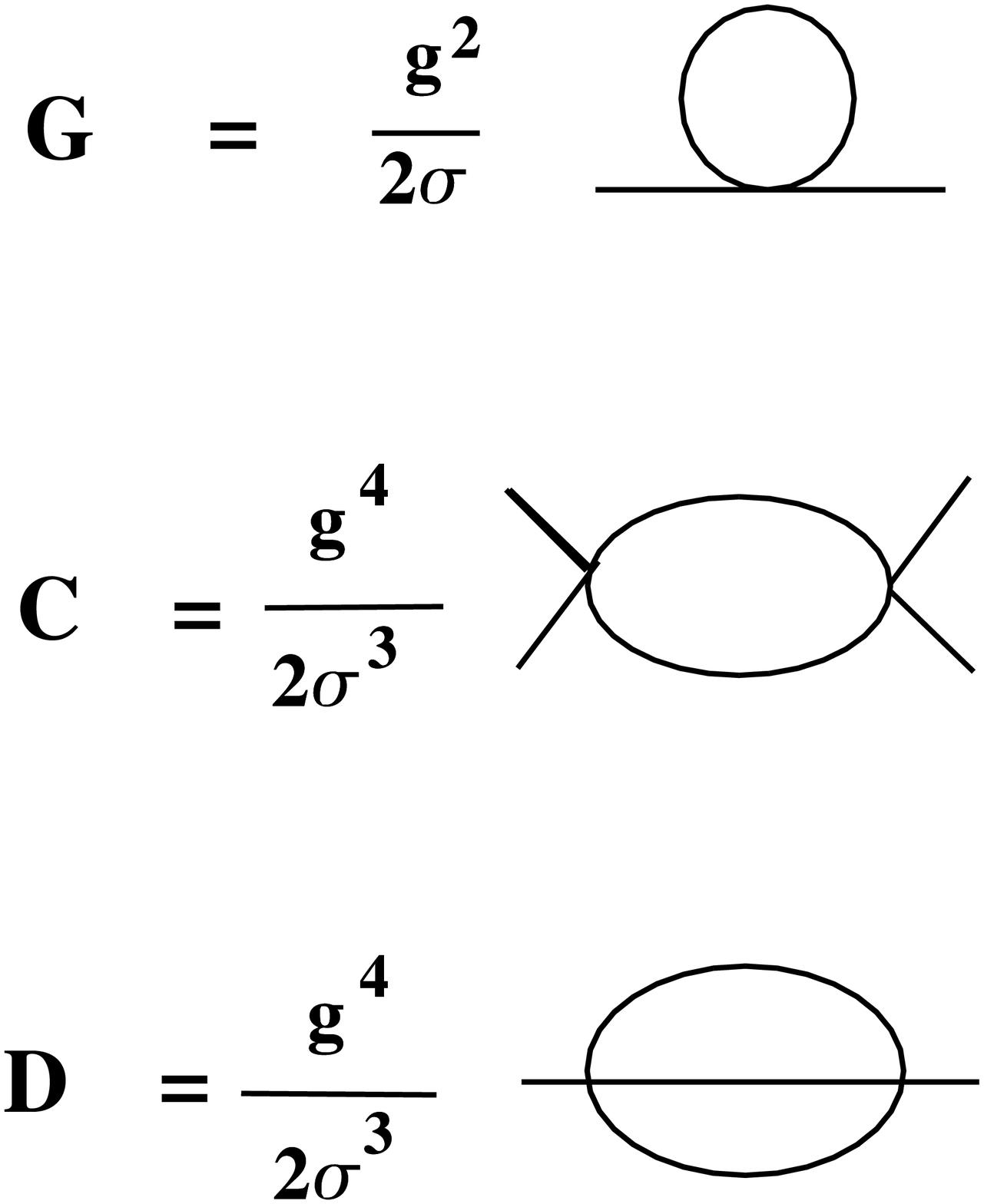}
\end{center}
\caption{\label{graf}
   }
\vspace*{-2mm}
\end{figure}
%%%%%%%%%%%%%%%%%%%%%%%%%%%%%%%%%%%%%%%%%%%%%%%%%%%%%%

%\vs
The full quantum field theoretical treatment of these factors is
rather involved.
In order to gain estimates over the parameters of a possible stationary
distribution of $A_0^2$ values, 
%-- and through this over the energy
%stored in the soft sector $E \approx V T^2 A_0^2 $, where $T$
%here is a typical hard momentum scale characteristic to the medium -- 
averages and correlations of $\gamma$ and $\xi$ have to be obtained.
This process can be facilitated by using a graphical notation:
external legs stand for the zero mode and internal lines for
the hard modes. (cf. Fig.\ref{graf}). 
For no spontaneous symmetry breaking, $F=0$ and $B=0$ immediately
follows, so the stationary distribution is a Tsallis-distribution
for $A_0^2$ and eventually in the soft energy 
$E \sim V\Lambda^2A_0^2 = A_0^2/(2m)$,
where $\Lambda$ is the underlying energy scale. 
A qualitative, order of magnitude estimate for the parameters of
the Tsallis distribution can be given as: $T \sim g^2\Lambda^4/(m\sigma^2)$
and $q - 1 \sim g^2\Lambda^2/\sigma^2$. The starting point of the
power law in eq.\eqref{TSALLIS} is at $E_c = T/(q-1)\sim\Lambda^2/m$.

%\vs
Finally we would like to check whether quantitative
estimates give reliable results. In high energy particle physics
experiments the transverse momentum distribution has been
investigated for long. From Gaussian fits to
the parton distribution\cite{PARTON-GAUSS} one conjectures a ratio 
$D/G = \exv{p_t^2} \approx 1\:  - \: 1.5$ GeV$^2$. On the
other hand power-law tails at high transverse momenta make a
value of $v \approx 5.8 \pm 0.5$ realistic. This fixes the
ratio to $G/C \approx 9.6 \pm 1$ 
and the Tsallis index to $q \approx 1.2 \pm 0.03 $.
The critical transverse momentum, beyond which the power-law
dominates the familiar Gauss distribution, can be calculated
from this to be $p_c \approx 3\: - \:4$ GeV. Compared to experimental
spectra\cite{POW-TAIL-SPECTRA} this is a quite realistic estimate.

\vspace{2mm}
%{\bf Conclusion.}
{\em In conclusion} we have shown that the smallest multiplicative
white-noise related to the classically deterministic damping constant
in the Langevin equation leads to a stationary distribution
of particle momenta, which ends in a power-law tail at high
values. This result seems to undermine the well-established
thermal approaches to phenomenological and field theoretical
studies in particle physics, where the presence of a multiplicative
noise is not less probable than the presence of an additive one
in any simplification (linearization) of the underlying
microdynamical problem. This approach on the other hand offers
a new way to deal with the interpretation of power-law tails
occurring in experimental findings, as well as it animates to seek
new methods in thermal field theory exceeding the traditional
thermodynamical approach. 
%stuck to the assumption of the Gibbs distribution.

%\vs
{\bf Acknowledgment}
Discussions with J. Zim\'anyi and A. Parvan about particle spectra
and thermodynamics are gratefully
acknowledged. Special thank from T.S.B. to B. M\"uller and C. Greiner with whom
he started to speculate about the role of noise in field theory.
This work has been supported by the
Hungarian National Research Fund OTKA (T034269, T037689, T034980, F043465).

%%%%%%%%%%%%%%%%%%%%%%%%%%%%%%%%%%%%%%%%%%%%%%%%%%%%%%%%%%%%%%%%%%

\end{document}